\documentclass[pra]{revtex4}
\usepackage{amsmath}
\usepackage{graphicx}
\usepackage{dcolumn}
\usepackage{bm}

\newcommand{\ket}[1]{| #1 \rangle}

\newcommand{\pscal}[2]{\langle #1 | #2 \rangle}
\begin{document}
\title[]{Optimized electron propagation on a quantum chain by a topological phase}
\author{Simone Paganelli}
\email{paganelli@ifae.es} \affiliation{Grup de F\'isica Te\`orica, Universitat
Aut\`onoma de Barcelona, 08193 Bellaterra (Barcelona), Spain} 
\author{Gian Luca Giorgi}
\affiliation{Institute for Cross-Disciplinary Physics and Complex Systems, IFISC (CSIC-UIB),
Campus Universitat Illes Balears, E-07122 Palma de Mallorca, Spain}
\author{Ferdinando  de Pasquale}
\affiliation{Dipartimento di Fisica, Universit\`{a} di Roma La
Sapienza, P. A. Moro 2, 00185 Rome, Italy}
\affiliation{INFM Center for Statistical Mechanics and Complexity}

\begin{abstract}
We study the quantum diffusion of an electron in a
quantum chain  starting from an initial 
state localized around a given site.
As the wavepacket diffuses, the probability of reconstructing the 
initial state on another site diminishes drastically 
with the distance.    
In order to optimize the state transmission 
we find that a 
topological quantum phase can be introduced.
The effect of this phase is the reduction of wavepacket spreading 
together with almost coherent group propagation. In this regime, the electron
has a quasi-linear dispersion and high fidelity can be achieved also 
over large distances in terms of lattice spacing. 
\end{abstract}
\maketitle                   





\section{Introduction}
The spatial  transmission of a quantum state is an important and 
nontrivial task in 
 quantum communication.
State encoding usually occurs on a local device and
an efficient channel is required to transmit it elsewhere. 
Quantum communication can be achieved by transmitting the particle 
on which the state has been 
encoded (flying qubit) \cite{cirac1}, by teleportation \cite{tele}  or by 
a quantum
state transfer in which only information is transmitted without transfer 
of matter or light. 
The 
first case can be, for example, a laser beam
that coherently propagates carrying the information 
codified on the polarization. Photons can travel 
with low loss, in optical fibres or even in free space, and can be
readily measured by a receiving party. This is because of the small interaction
with the environment and the linear dispersion, allowing for a 
propagation without spreading of the wavepacket.
This is why optical flying qubits 
represent a very efficient channel over long distances.

Nevertheless, different solutions  can be  more suitable over smaller
(micro and nanometre) distances typical of solid state devices.
To this aim, an ion trap based device has been proposed\cite{kielpinski-short,blatt2008}.
Other schemes have been
also described for short distance communication by a spin chain
used as channel (see \cite{ammsagato,bose,subry, subrahmanyam-entanglement1,ammbugarth,ammrossini2}
and references therein).
Here, the transfer is based on the diffusion through the 
chain of an initially localized spin state by means
of typical collective modes induced by the particular phase in which we prepare it. 
The use of local excitations requires an optimization of the interference 
for the state reconstruction\cite{osborne,giampaolo1,burgath,datta,gualdi:022325,ammbugarth1}, 
 because there is, 
together with the usual problem of decoherence, also a diffusion phenomenon due to
 typically non-linear dispersion laws.
By engineering the couplings of the spin-chain  and local end-chain operations, a perfect state transfer can be achieved \cite{difrancoPRL08}.
Other different physical realizations of quantum channels have been also
suggested with different strategies to optimize the state transfer:  Josephson
arrays \cite{romito}, nanoelectromechanical oscillators
\cite{eisert}, quantum chains as quantum bus \cite{quantumbus,plenio-anello,wojcik,giorgi:153308}, spin-$1$
chains\cite{oriol2007NJPh,romero-isart:050303}, dot chain \cite{noi,0312112,ammliu,ammliu1,ammwang}. 

In this paper, we propose a scheme of quantum state transfer over a
quantum chain in which an electron, and the spin state encoded on it,
 plays the role of flying qubit.
We study the diffusion of an electron in a tight-binding
chain, starting from an initial localized state, under the action 
of a topological phase, induced, for example, by a magnetic   
Aharonov-Bohm flux \cite{eherenberg-siday,A-B}.
The electron wavepacket propagation in chain of quantum 
dots was studied in  \cite{nikolopoulos2004},
while a general treatment of the Aharonov-Bohm scattering for a free particle was given by Stelitano 
\cite{PhysRevD.51.5876}. In Ref. \cite{ciccarello2007}, it has been shown that the Aharonov-Bohm effect enables the generation of entanglement in mesoscopic rings.

We derive the dynamics of the electron in a discrete one-dimensional lattice, showing how  
a suitable topological phase can induce an almost linear dispersion. As a result, 
in the limit of a large number of sites, the electron moves  coherently with a 
reduced  wavepacket spreading.
Such a scheme allows for communication over an intermediate range, contrary to what happens
with other dispersive channels. A different use of the topological phase
was used in \cite{giampytopo} for communication on spin rings.

We show that an initial amount of delocalization is necessary for the scheme to work properly, otherwise
an initial completely localized state does not feel the effect of the topological phase, spreading rapidly.  
The transmission efficiency can be characterized by the fidelity
\cite{nielsen-qcomputer,schumacher-transmission}. Even if the
definition of this quantity is more general, for a pure state 
it reduces to the squared modulus of the
projection of the evolved state on the transmitted one.
It quantifies how much the initial state, centered around the $0-$site, can be reproduced 
around another  given site of the chain after a suitable time.
The other quantity we are going to study is the  evolution of the probability distribution, to better
emphasize the two aspects of the evolution: the propagation, due to the presence of a group velocity
over different chain modes, and the wavepacket spreading due to different phase velocities. 

In the next section we describe the system and calculate exactly both the fidelity and the probability distribution
in the general case.
In the section III, we show how the localized preparation is not useful for communication. In section IV
we study the evolution of an initial square packet, comparing the result with a simple analytical 
approximation, and show how a good communication can be achieved over a large number of sites.
Finally we give our conclusions.

\section{Model}
Let us consider a single electron moving in a one-dimensional ring-shaped lattice, 
with $N$ sites and lattice constant $a$.
In the reciprocal lattice space, and in the tight-binding approximation, 
the diagonal Hamiltonian  is
\begin{equation}\label{eqn:hringdot}
    H_R=\sum_{q,\sigma} \epsilon_q c^\dag_{q,\sigma}  c_{q,\sigma},
\end{equation}
where $\epsilon_q=-2w \cos (a q)$ are the energies forming the
lattice band, $w$ is the half band width,  $q=\frac{2 \pi}{N} n$
are the vectors of the reciprocal lattice and $c_{q,\sigma}$ the
electron fermionic operators. 
It is simple to see that the operators time dependence is given by $c^\dag_{q,\sigma}(t)=e^{-i
\epsilon_q t} c^\dag_{q,\sigma}$. Hereafter, we shall impose $\hbar=1$.

In a perspective of quantum communication, we can encode the qubit
on the spin state since this latter does not change during the evolution. From now on,
we shall omit the $\sigma$ index, denoting with 
$c^\dag=\alpha_1 c^\dag_\uparrow+\alpha_2 c^\dag_\downarrow$ the electron operator with generic
spin state.

We start  preparing the electron in a state localized around the site $j=0$
\begin{equation}
    \ket{\psi_{0}(0)}=\sum_j g_j c^\dag_{j}\ket{0}
    =\frac{1}{\sqrt{N}} \sum_q \tilde{g}_q
    c^\dag_{q}\ket{0},
\end{equation}
where $g_l$ is an amplitude distribution centered around the site $j=0$
and 
\begin{equation}
    \tilde{g}_q= \sum_j g_j e^{ i q a j},
\end{equation}
is its Fourier transform. The probability distribution of the site occupancy is 
\begin{equation}
P_j(0)=\left|\langle0|c_{j}|\psi_{0}(0)\rangle\right|^{2}=\left| g_j	\right|^2.
\end{equation}
As soon as the electron evolves, a diffusion process starts. The state evolution is easily calculated
\begin{equation}
    \ket{\psi_{0}(t)}=\frac{1}{\sqrt{N}}\sum_q \tilde{g}_q e^{-i
\epsilon_q t} c^\dag_{q} \ket{0},
\end{equation}
so as  the time-dependent probability distribution 
\begin{equation}\label{eqn:Prob}
P_{j}(t)=\left|\langle0|c_{j}|\psi_{0}(t)\rangle\right|^{2}=\frac{1}{N^{2}}\sum_{qq'}\tilde{g}_{q}\tilde{g}_{q'}^{*}e^{-i\left[(\epsilon_{q}-\epsilon_{q'})t-(q-q')aj\right]}.
\end{equation}

As expected, since $\epsilon_q=\epsilon_{-q}$, one can see from Eq. (\ref{eqn:Prob}) that $P_{j}(t)$  remains centered around $j=0$, and a pure diffusion occurs. In order to obtain a real transport of the wave packet, it is necessary to introduce something breaking the translational invariance of the Hamiltonian.
This result can be achieved introducing a topological phase which changes the hopping terms 
as follows: $c^\dag_j c_{j+1} \rightarrow e^{i \theta} c^\dag_j c_{j+1}$.
As a consequence, the energies become
\begin{equation}
\epsilon_q(\theta)=-2w \cos\left(a q-\theta  \right).
\end{equation}

As we shall show, tuning the $\theta$ phase makes it possible to reduce 
the wave packet spread and optimize the particle transmission.
Such a phase can be obtained, for instance, introducing an suitable magnetic
orthogonal field $B$ trough the ring. By means of the
Aharonov-Bohm effect the   phase shift is 
$\theta =(2 \pi \Phi)/(N \Phi_0)$ where $\Phi$ is the magnetic flux
through the ring and $\Phi_0=hc/e$ is the quantum unit of
magnetic flux. 
Topological phases can be also created in different ways, for example by means of the electric
Aharonov-Bohm effect \cite{elettro1998Natur.391..768V,1999JAP....85.1984F,1993PhRvB..48.1537T}. We will not
insist on this point, referring to a generic phase independently on how to generate it.

In a quantum information context, where the conducting ring can be  used as a 
communication channel (quantum bus), it is useful to consider another quantity,
the so-called fidelity. It represents how the initial state, centered on the $j=0$ site, can be
efficiently reproduced around another site $j=d$ (the \emph{receiver} site). Starting from the transfer 
amplitude
\begin{equation}\label{eqn:fd}
f_d(t)=\pscal{\psi_{d}}{\psi_{0}(t)}=\frac{1}{N}\sum_q |\tilde{g}_q|^2 e^{i (q a d-\epsilon_q t)},
\end{equation}
the fidelity is defined as
\begin{equation}
F_d(t)=\left| f_d(t) \right|^2 .
\end{equation}
In the following sections we shall examine two cases: in the first one, the electron is initially localized on the
site $j=0$; in the other case, the electron is prepared with a squared distribution around the same site. 

\section{Atomic preparation }
To start with,  we study the case in which the electron is prepared on one site, say $j=0$,  putting 
$g_j=\delta_{j,0}$ and $\tilde{g}_q=1$. We shall refer to this preparation as the \emph{atomic} one.
The transfer amplitude is simply

\begin{equation}\label{eqn:effe}
    f_d(t)=    \frac{1}{N}\sum_q e^{i (q a d-\epsilon_q t)}.
\end{equation}
In this case, the probability distribution is equal to the fidelity  $  P_{j}(t)=F_j(t) $.
In the limit of large number of sites, $f_d$ is proportional to a
Bessel Function \cite{abram} $J_d$
\begin{equation}
    f_d(t) = e^{i \frac{\pi}{2} d}  J_{d}( 2 w t).
\end{equation}

\begin{figure}[htbp]
\begin{center}
\includegraphics[height=8cm,angle=-90]{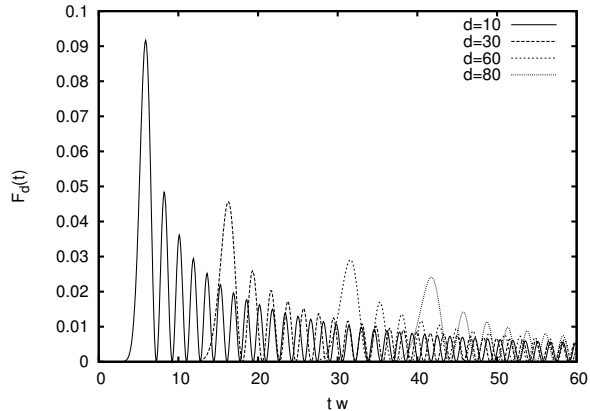}
\end{center}
\caption{Time evolution of the fidelity in the atomic limit with fixed 
final site $d$ and $N=500$ . Fidelity is plotted for $d=10,30,60,80$.  }
\label{fig:fid00}
\end{figure}

For fixed $d$, the fidelity has a maximum for $t\simeq d/2w$, which is 
also an absolute maximum in the $N \rightarrow \infty$  limit.
This can be seen  in Fig. \ref{fig:fid00} where the time dependence of the fidelity
is plotted for different final sites. For finite value of $N$,  
other maxima appear, higher than the first,  because of  the
interference between the two different propagating wave packet in which
the initial wave function is split. This effect occurs also  in the context of
quantum communication on spin chains\cite{bose}. Here, 
the interference is due to the counterpropagating spin waves and the 
constructive peaks are used to optimize the short range transport (few sites).
This mechanism is not very interesting on larger distances because of the
difficulty of estimating the optimal times of the better constructive
interference and  their strong dependence on  the site  number. For this
reason we work in the limit of large number of sites, fixing the
first maximum time as the useful one for the transport process.

The atomic preparation appears to be highly inefficient, as it can be 
seen looking at the very small fidelities in Fig. \ref{fig:fid00}. Moreover,
is easy to demonstrate that, in the large $N$ limit, the introduction of a  topological phase 
does not change anything (changing $aq-\theta$ into $a q'$ introduces,  in
the continuum limit of Eq. (\ref{eqn:effe}), only a inessential phase shift). 
As we shall discuss in the next section, fidelity increases  if the electron wavefunction
is initially delocalized over few sites. 

\section{Square packet preparation}
The second configuration we consider 
consists in an electron prepared  to 
be equally delocalized  around the site $j=0$.
The initial state is given by 
a square wave packet centered on the $0$
site and extended over $2M+1$ sites. 
The amplitude distribution is
\begin{equation}
    g_l= \left\{\begin{array}{cc}
      \frac{1}{\sqrt{2M+1}}   & \mbox{      If $ -M \leq l \leq M$} \\
      0                       & \mbox{   Elsewhere}, \\
    \end{array}
    \right.
\end{equation}
corresponding to the form factor
\begin{equation}\label{eqn:gtilde}
    \tilde{g}_q=\frac{1}{\sqrt{\lambda}}
    \frac{\sin{\lambda \frac{q a}{2}}}{\sin \frac{q a}{2}},
\end{equation}
with $\lambda=(2M+1)$. 

\begin{figure}[htbp]
\begin{center}
\includegraphics[height=8cm,angle=-90]{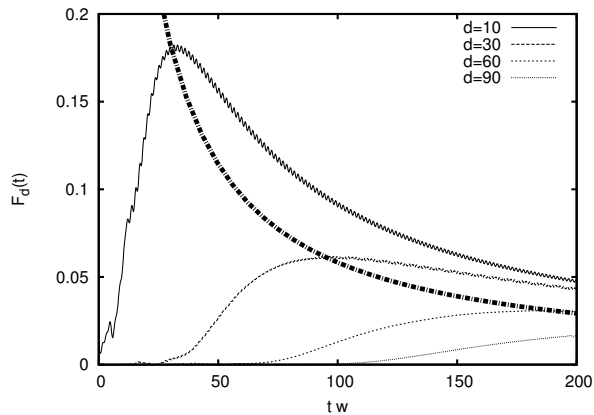}
\end{center}
\caption{Time evolution of the fidelity in the square packet preparation ($M=5$), without phase $\theta$,  with fixed 
final site $d$ and $N=500$ . Fidelity is plotted for $d=10,30,60,90$.
Thick line indicate the maxima of fidelity reached by each site at different times.}
\label{fig:fid02}
\end{figure}

\begin{figure}[htbp]
\begin{center}
\includegraphics[height=8cm,angle=-90]{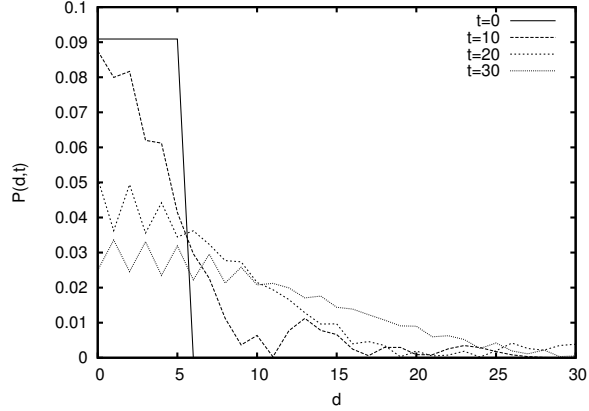}
\end{center}
\caption{Probability distribution for $N=500$, in the square packet preparation ($M=5$) and without phase $\theta$. The function is symmetric and here is 
plotted only for $d$ positive. The wavepacket spread is shown for different times.
$t=0,10,20,30$. }
\label{fig:distr02}
\end{figure}

The fidelity is plotted in Fig. \ref{fig:fid02} for a 
$500$-site chain and without any phase $\theta$. 
One can see that a completely different behavior occurs,  even if
the efficiency still remains low. The reason is to 
be ascribed to the fact that the wave packet actually does not 
moves but only diffuses, as it is show in Fig \ref{fig:distr02}, 
where the probability distribution is reported for different
times.

The introduction of a suitable phase $\theta$ induces a wavepacket propagation
so to increase the transmission efficiency. In order to estimate the optimal 
value of the phase, we derive here an approximate analytical expression 
for $F_d(t)$ and $P_d(t)$. Considering a very large $N$ we can
substitute the sum (\ref{eqn:fd}) by an integral introducing the continuous
variable $x=q a/2$. The expression for the transfer amplitude, with a 
generic $\theta$ becomes
\begin{equation}\label{eqn:int}
    f_d(t)\simeq \frac{1}{\pi }
    \int^{\frac{\pi}{2}}_{-\frac{\pi}{2}}dx
    G(x)
    e^{i(2xd+2w t\cos{(2x - \theta)})}.
\end{equation}
The form factor $G(x)=\lambda^{-1} \left(\sin{\lambda x}/ \sin x  \right)^2$ 
in the integral is a periodic function 
whose peaked principal maxima are spaced by an amount of $\pi$ from each other. The integration interval
contains only the principal maximum centered on the origin and so
we can expand $G(x)$ around $x=0$ approximating it by a Gaussian
function
\begin{equation}\label{eqn:forma}
    G(x) \simeq
    \lambda \left(1- \frac{\lambda^2-1}{3}x^2\right)\simeq
    \lambda e^{-\frac{\lambda^2-1}{3}x^2}.
\end{equation}
Now we expand the argument of the phase factor into the integral.
We note that the expansion is to be performed, at least, until the
second order because the first term vanishes as soon as the
phase is turned off. So, if we want to take into account
of the effect of phase-free propagation we have to
consider the expansion
\begin{equation}\label{eqn:coseno}
    \cos{(2x - \theta)}\simeq\cos{\theta}+2\sin{\theta}x-2 \cos{\theta}x^2.
\end{equation}
Isolating the inessential phase factors we obtain
\begin{equation}\label{eqn:app1}
    f(t)\simeq e^{i\phi} \frac{\lambda}{\pi}\int^\infty_{-\infty}dx
    e^{-\left[\frac{\lambda^2-1}{3}-i 4 w t \cos{\theta}\right]x^2+
    i 2 x (d+2wt \sin\theta)},
\end{equation}
that can be integrated as a common Gaussian integral. The resulting fidelity
is
\begin{equation}\label{eqn:fid}
    F_d(t)=A(t)
    e^{-\frac{
    \left[d+2wt \sin\theta\right]^2}{2 \sigma_F^2(t)}},
\end{equation}
with 
\begin{equation}\label{eqn:amp}
    A(t)=\frac{3 \lambda^2}{\pi\sqrt{(\lambda^2-1)^2+144 w^2 t^2 \cos^2 \theta}},
\end{equation}
and 
\begin{equation}\label{eqn:var}
    \sigma_F^2(t)=  \frac{(\lambda^2-1)^2+144 w^2 t^2 \cos^2 \theta}{12 (\lambda^2-1)}. 
\end{equation}

The same calculation can be done for the probability distribution. 
Expanding  
the form factor (\ref{eqn:gtilde})  into 
\begin{equation}
\tilde{g}(x)\simeq \sqrt{\lambda} e^{-\frac{\lambda^{2}-1}{6}x^{2}},
\end{equation}
the wave function   (\ref{eqn:Prob}) becomes
\begin{equation}
\psi_{d}(t)=\simeq 
e^{i\phi} \frac{\sqrt{\lambda}}{\pi}\int^\infty_{-\infty}dx
e^{-\left[\frac{\lambda^2-1}{6}-i 4 w t \cos{\theta}\right]x^2+
    i 2 x (d+2wt \sin\theta)}.
\end{equation}
Integrating and squaring, we obtain 
\begin{equation}\label{eqn:appP}
P_{l}(t)=B(t)
    e^{-\frac{
    \left[d+2wt \sin\theta\right]^2}{2 \sigma_P^2(t)}},
\end{equation}
with 
\begin{equation}
B(t)=\frac{6 \lambda}
{\pi\sqrt{(\lambda^{2}-1)^{2}+576w^{2}t^{2}\cos^{2}
\theta}}
\end{equation}
\begin{equation}
\sigma_P^2(t)=
\frac{(\lambda^{2}-1)^{2}+576w^{2}t^{2}\cos^{2}\theta}{24(\lambda^{2}-1)}.
\end{equation}

By this calculation, both $F_d(t)$ and $P_d(t)$  are approximated 
with Gaussian functions of the variable $d$,  which propagate and diffuses in time. 
The propagation velocity
is given by $v=-2w \sin \theta$, while the diffusion is quantified by the variance
$\sigma_{F,P}^2$. We shall study $F_d(t)$ as a function of time with fixed $d$ 
(the \emph{receiver} site),  and $P_d(x)$ as a spatial distribution at fixed times.
Two different behaviors  appear, corresponding to $\theta=0$ and $\theta=-\pi/2$.

In the first case, with $\theta=0$,   the quadratic term in the
expansion (\ref{eqn:coseno}) prevails. There is no propagation of the probability 
distribution, $v=0$. The wave packet diffuses only, as one can see in Fig. \ref{fig:distr02}.   The fidelity has a maximum in correspondence of the time  
\begin{equation}\label{eqn:tmax1}
t^\ast = \frac{\sqrt{ (\lambda^2-1)(12 d^2-\lambda^2+1)}}{12 w}.
\end{equation}
In Fig. \ref{fig:fid02} is depicted the fidelity at different sites with 
the curve of the maxima. The result is better than the atomic case but the efficiency rapidly 
decreases with the distance.

\begin{figure}[htbp]
\begin{center}
\includegraphics[height=8cm,angle=-90]{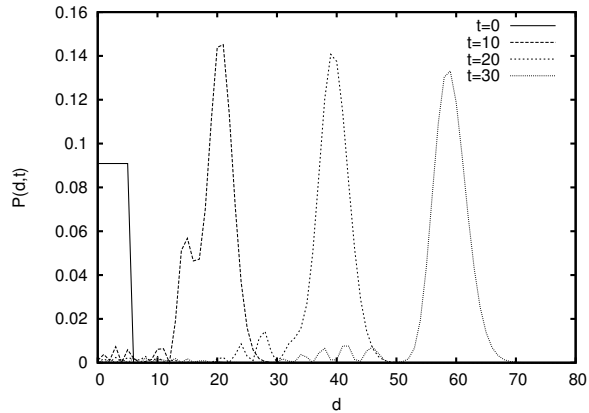}
\end{center}
\caption{Probability distribution for $N=500$, in the square packet preparation ($M=5$) and with phase $\theta=-\pi/2$. Here a propagation of the wavepacket can be observed.  
The wavepacket spread is shown for different times.
$t=0,10,20,30$.}
\label{fig:distr01}
\end{figure}

\begin{figure}[htbp]
\begin{center}
\includegraphics[height=8cm,angle=-90]{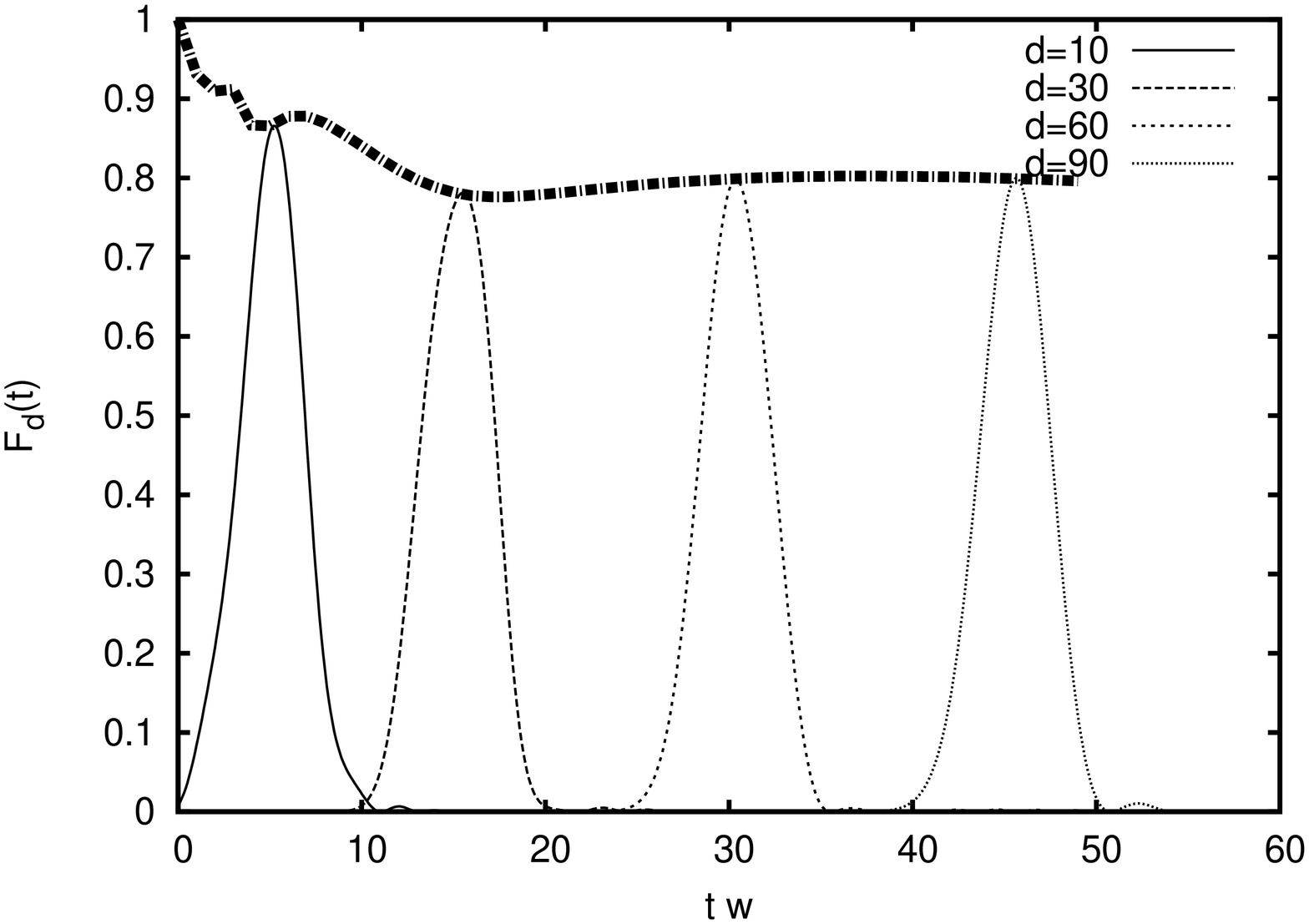}
\end{center}
\caption{Time evolution of the fidelity in the square packet preparation ($M=5$), with phase $\theta=-\pi/2$,  with fixed final site $d$ and $N=500$ . Fidelity is plotted for $d=10,30,60,90$.
The thick line indicate the maxima of fidelity reached by each site at different times.}
\label{fig:fid01}
\end{figure}

\begin{figure}[htbp]
\begin{center}
\includegraphics[height=8cm,angle=-90]{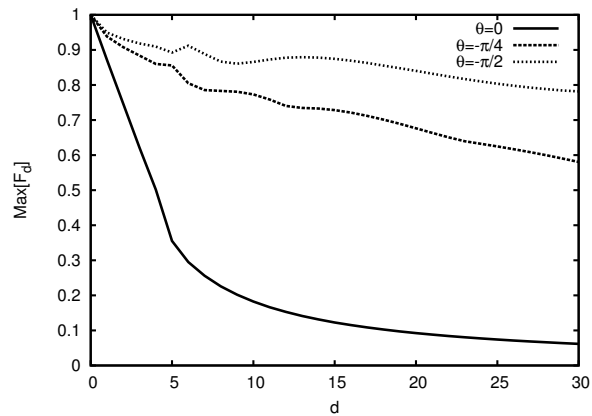}
\end{center}
\caption{Maximum of fidelity with distance.$N=500$ and $\theta=0,-\pi/4,-\pi/2$. 
The topological phase allows an efficient long range state transfer.}
\label{fig:maxima}
\end{figure}

In the second case, with $\theta=-\pi/2$, only the linear term in (\ref{eqn:coseno})
remain. The dispersion becomes approximately linear and the  wavepacket (\ref{eqn:appP})) 
does not diffuse (Fig. \ref{fig:distr01}). Moreover, it propagates with velocity  $v=2w$  causing an enhancement in the fidelity. In particular, $F_d(t)$ assumes its maximum value at the approximate time
\begin{equation}\label{eqn:tmax2}
   t^\ast = \frac{d}{2 w}.
\end{equation}
 In  Fig. \ref{fig:fid01} time dependent fidelity is reported for different 
receiving sites, together with the time evolution of the maxima. As one can see, 
in this case a fidelity of the order of about $0.8$ is achieved  even for 
distances of the order of $100$ sites. Notice that the maximum value of the fidelity one can achieve is almost independent of the chain length.
In Fig. \ref{fig:maxima} the maximum of fidelity is reported as a function of distance for
the two extremal values of $\theta$  and an intermediate one.

Other preparations, even
more experimentally feasible, can be introduced as, for example, a Gaussian distribution.
In this case, being our analytical treatment basically a Gaussian approximation, we
expect an even better fitting with the numerical results.
The calculations for every type of initial distribution,
can be done in the same way introducing a suitable $\tilde{g}_q$ in Eq. (\ref{eqn:fd}). 
Nevertheless, it is enough to show the simple square-packet case to 
meet all the interesting physical effects leading to  coherent propagation.

\section{Conclusions}
We studied the quantum diffusion of an electron
in a periodic lattice, in the tight-binding 
regime. We shown that,even in the presence of a nonlinear 
dispersion,  it is possible to approach a linear 
regime where the electron wave packet spreading is reduced.
This is possible by introducing in the system a suitable
topological quantum phase.   
The new advances in the scalable quantum 
information and communication on mesoscopic
solid state devices  gives rise  to a 
need for new communication channel beyond the usual photon flying qubit.
In this perspective, the possibility of a flying qubit  
carried by electrons appears a promising resource for state transfer along 
mesoscopic scales. . 

\section{Acknowledgment}
The authors wish to acknowledge  EU IP SCALA and  Spanish MEC grant FIS 2008-/01236
QUICSUAP. GLG acknowledges the “Juan de la Cierva” fellowship of the Spanish Ministry of Education.

%
\bibliographystyle{fdp}
\bibliography{/media/disk/universita/bibliografia/bibl-cont_var,/media/disk/universita/bibliografia/topological,/media/disk/universita/bibliografia/bibliografia,/media/disk/universita/bibliografia/bibl-bosehubbard,/media/disk/universita/bibliografia/bibl-iontraps} 

\end{document}